# Defect Management Using Depth of Inspection and the Inspection Performance Metric

**T.R. Gopalakrishnan Nair,** Aramco Endowed Chair-Technology, PMU, KSA
**V. Suma,** RIIC, Dayananda Sagar Institutions

**Abstract.** Advancement in fundamental engineering aspects of software development enables IT enterprises to develop a more cost effective and better quality product through aptly organized defect management strategies. Inspection continues to be the most effective and efficient technique of defect management. To have an appropriate measurement of the inspection process, the process metric, Depth of Inspection (DI) and the people metric, Inspection Performance Metric (IPM) are introduced. The introduction of these pair of metrics can yield valuable information from a company in relation to the inspection process.

### Introduction

A defect in software is expensive especially when it dwells and manifests. One of the prevailing challenges in the software industry is therefore the production of defect-free software [1]. The continuance of IT enterprise, hence, depends upon the choice of apt defect management strategies in order to generate defect-free software.

Quality control and quality assurance techniques are the two most successful defect management strategies. Quality control activity is for the product and quality assurance techniques are for the process. The current trends in the industry concentrate on testing, which is a quality control activity. However, what matters is the process through which a product is developed, and therefore excellence in the process plays a vital role towards delivery of high quality products. Among several techniques applied for quality assurance in the software field, like walkthrough, inspection, and review, inspection is one of the promising techniques for defect management. Despite its perceptible significance, inspection is either very casually treated or more often overlooked and many times it is maintained only for accounting purposes. One of the rationales being projected to escape the vital process step is identifying it as a mind-numbing, lengthy activity rather than a quality improvement process.

Since quality is a quantifiable unit, this article aims to draw the attention of the software community including management, developing teams, stakeholders, and outsourcing agents to an important aspect of bringing in a cultural change. It is worth for the industry to notice and comprehend the connotation of software inspection as process integration introduced recently by the implementation of a pair of metrics that are meant to measure the quality level of inspection process and further measure the competency of the people. The two metrics are DI, a process metric, and IPM, a people metric [2].

### The DI Metric

Let us have a closer look at the apparently simple but powerful concept of DI. Let $N_i$ be the number of defects captured by the inspection process and $T_d$ be the number of defects captured by both inspection and testing approaches.

$$DI = N_i / T_d$$

*Equation A:*

DI can be measured phase-wise or before the deployment of the product using the above metric.

DI evaluation is realized in two phases. In the first phase, DI is calculated using shop floor defect count for a particular set of projects. This phase enables the software company to analyze the depth in which inspection process has occurred for a set of particular projects either phase-wise or at the project level. From our deep and rigorous investigations carried out across several service-based and product-based software industries of various production capabilities, it is found that the DI value varies from project to project. The metric is distinctive as it quantifies the inspection process with measurable levels, which is not observed in current industry standards. DI is considered to be in the range of zero to one where zero is nil performance and one indicates 100% defects captured exclusively through inspection process, which is hardly ever possible. An inspection level of 0.3 to 0.5 is considered normal inspection process and a level of 0.5 onwards requires high competency in the process [2].

Table 1 specifies the ranges of DI values [2]. With this chart, it is now possible for software personnel and all stakeholders to identify the maturity level of the company and enable them to either continue with the existing level or formulate strategies towards up gradation of their level. An additional strength of this mode of quality measurement is to throw light towards prediction of desired level of inspection.

| DI | Inspection performance |
|---|---|
| 0 - 0.1 | Worse (W) |
| 0.1 – 0.2 | Very Low (VL) |
| 0.2 – 0.3 | Low (L) |
| 0.3 – 0.4 | Normal (N) |
| 0.4 – 0.5 | Above Normal (AN) |
| 0.5 – 0.6 | High (H) |
| 0.6 – 0.7 | Very High (AV) |
| 0.7 – 0.8 | Best (B) |
| 0.8 – 0.9 | Excellent (E) |
| 0.9 - 1 | Ideal (I) |

*Table 1: Range of DI Values*

Prediction of quality of inspection process is not yet achieved. DI is a process metric that can predict the quality of software inspection process. The second phase consists of prediction of DI value for a new project through the approach of mathematical modeling scheme, which uses Multiple Linear Regression (MLR) models.

Prediction of DI is through the evaluation of process coefficients from the historical projects. Process coefficients are a set of β constants ($β_0$ to $β_4$) which are evaluated using least square estimates or using Matlab support. A minimum of five projects ($P_1$ to $P_5$) is required to evaluate the process coefficients. However, at a larger scale, depending on the history of the company and the past records of the projects that the company had handled, several groups of samples can be taken. It is also observed from the investigation made by several researches that effectiveness of inspection is influenced by four major and mutually exclusive parameters [2]. They are:

$x_1$ = inspection time
$x_2$ = preparation time
$x_3$ = number of inspectors
$x_4$ = experience level of inspectors

Having obtained the process coefficients and substituting desirable values to the inspection influencing parameters, it is now possible for the software company to predict the value of dependent parameter Y (DI) as given in equation (B) [3].

$$Y = β_0 + β_1x_1 + β_2x_2 + β_3x_3 + β_4x_4 + e$$

Equation B:

Matrix representation for the prediction of DI is given in equation (C).

Software companies using the stabilized process coefficients can now predict the desired level of DI for any project ($P_i$) by modulating the inspection influencing parameters. Alternatively, with the evaluation of process coefficients it is also possible to tune the values of the inspection influencing parameters to achieve the desirable DI value. Table 2 illustrates the DI computation of 15 projects that are sampled from various product-based and service-based software industries. The sampled projects depicted throughout this article are similar types of projects that are developed using Java and operate in a similar type of environment.

Discernible benefits of DI in software organization are:

1. DI is a quality metric introduced in order to quantify the depth in which the inspection process is performed.
2. The objective of introducing DI as a defect detection metric is to enable one to analyze the defect capturing ability of the company through an inspection approach.
3. DI is a defect preventative metric whose implementation in the software industry acts as a lesson learned from previous projects with regard to the depth in which inspection is conducted and thus indicates the inspection team either "to improve the inspection performance" or "to maintain the desired level of inspection."
4. The aim of DI as an indicator metric is to inform the test team of the depth in which defects are detected through the

$$[DI] = [X][β] + [E] \text{ where}$$

$$[DI] = [Parameters] \times [processCoefficients] + [Error\ Term]$$

$$DI = \begin{bmatrix} DI_1 \\ DI_2 \\ DI_3 \\ DI_4 \\ - \\ - \\ DI_n \end{bmatrix} \quad β = \begin{bmatrix} β_0 \\ β_1 \\ β_2 \\ β_3 \\ β_4 \end{bmatrix} \quad X = \begin{bmatrix} 1 & x_{11} & x_{12} & x_{13} & x_{14} \\ 1 & x_{21} & x_{22} & x_{23} & x_{24} \\ 1 & x_{31} & x_{32} & x_{33} & x_{34} \\ 1 & x_{41} & x_{42} & x_{43} & x_{44} \\ - & - & - & - & - \\ - & - & - & - & - \\ 1 & x_{n1} & x_{n2} & x_{n3} & x_{n4} \end{bmatrix} \quad E = \begin{bmatrix} e_1 \\ e_2 \\ e_3 \\ e_4 \\ - \\ - \\ e_n \end{bmatrix}$$

Equation C:

|  | P1 | P2 | P3 | P4 | P5 | P6 | P7 | P8 | P9 | P10 | P11 | P12 | P13 | P14 | P15 |
|---|---|---|---|---|---|---|---|---|---|---|---|---|---|---|---|
| Project hours(*) | 250 | 263 | 300 | 507 | 869 | 1806 | 2110 | 4248 | 4586 | 4644 | 6944 | 7087 | 7416 | 8940 | 9220 |
| DI at req. phase | 0.53 | 0.49 | 0.67 | 0.52 | 0.33 | 0.48 | 0.5 | 0.46 | 0.39 | 0.27 | 0.44 | 0.44 | 0.49 | 0.44 | 0.47 |
| DI at des. phase | 0.5 | 0.38 | 0.46 | 0.54 | 0.42 | 0.5 | 0.44 | 0.49 | 0.44 | 0.4 | 0.64 | 0.46 | 0.57 | 0.45 | 0.43 |
| DI at imp. phase | 0.5 | 0.57 | 0.44 | 0.53 | 0.37 | 0.21 | 0.39 | 0.51 | 0.51 | 0.4 | 0.55 | 0.5 | 0.46 | 0.47 | 0.49 |
| Avg DI | 0.51 | 0.48 | 0.52 | 0.53 | 0.37 | 0.40 | 0.44 | 0.49 | 0.45 | 0.36 | 0.54 | 0.47 | 0.51 | 0.45 | 0.46 |
| Tc (%) | 96.0 | 95.0 | 91.5 | 96.0 | 89.8 | 87.0 | 92.0 | 95.4 | 96.5 | 88.3 | 96.9 | 96.5 | 93.1 | 95.8 | 92.3 |

(*) Total project time is measured in terms of person hours and contains documentation times, training time and release time etc., which are not relevant for this discussion; P = Project; req – Requirements analysis phase; des – Design phase; imp – Implementation phase; Avg - Average; Tc –Total defects captured in the complete project

Table 2: DI Estimation

inspection approach. Thereby, it directs the test team to frame effective strategies to eliminate remaining defects.

5. The rationale for the introduction of DI is to provide a deep visibility to the inspection team, the company management, outsourcing agents, and other stakeholders about the depth in which inspection is performed and thereby to provide enough transparency in the process.

6. It is observed that CMMI® Level 4 and above certified software industries are capable of performing inspection with an average DI value of 0.4 to 0.5. This demands that the test team puts in a certain amount of test effort to detect and eliminate remaining defects. A decrease in DI value further demands increased test effort, test time, increased developmental cost, and rework cost.

7. Project success depends upon quality [4]. Business success depends upon cost of quality [5]. According to the cost quality analysis, cost of rework is usually several orders higher at final stages when compared to quality implementation at initial stages [6]. The variation depends on the phase in which a defect originated and was later detected. The philosophy and culture that we propagate here offers a watertight control over defect management with appropriate quantitative metrics and methods in the process domain. Hence, the DI metric, with its distinguishing feature of giving process visibility, paves the way for stakeholders to control their developmental cost.

8. Implementation of DI is therefore a billion dollar savings to a software company since they are now able to visualize the depth in which all static defects are recovered and hence plan only towards detection and removal of dynamic defects through testing activities.

9. The existence of a DI metric is an eye opener for all stakeholders including clients and outsourcing agents to justify and control the developmental cost. With this metric, they are therefore in a position to substantiate the quality of the product and the maturity level of the company dynamically from project to project and predict a price tag for their project.

10. The DI metric brings out the variation in quality level of functional inspection and predicted inspection. This knowledge further enables the developing team to equip themselves towards their augmentation activities in order to endure in the competitive atmosphere of software industry.

### The IPM Metric

Effectiveness of the inspection process depends on the people who drive the process. However, there are no software quality metrics existing to measure the performance of the inspection team within the constraints of major inspection affecting parameters, namely 1) inspection time, 2) inspection preparation time, 3) number of inspectors, 4) experience level of inspectors, and 5) complexity of the project that is measured in terms of function points [2]. The IPM metric helps a software company to make decisions toward the selection of appropriate values to the aforementioned parameters subsequently opting for the desirable team performance.

$$IPM = N_i / IE$$
$$\text{where } IE = N \times T$$
$$\text{and } T = I_t + P_t$$

*Equation D:*

Let $N_i$ = Number of defects captured by inspection process and IE = Inspection Effort

Where IE = Total number of inspectors (N) × Total amount of inspection time (T)

Total amount of inspection time (T) = Actual inspection time ($I_t$) + Preparation time ($P_t$), (taken per person)

Where IE = Total number of inspectors (N) × Total amount of inspection time (T)

Total amount of inspection time (T) = Actual inspection time ($I_t$) + Preparation time ($P_t$)

IPM can be realized in two stages. In the first stage, number of defects captured by the inspection team within the aforementioned parameter constraints for any particular project is found using shop floor defect count. This mode of IPM calculation enables the software team to measure the team performance properties.

The second stage of realizing IPM is to predict IPM value for a new project using a mathematical scheme. Prediction of IPM for a project is realized using MLR models.

Let ($\beta_0$ to $\beta_5$) = team coefficients,
 $x_1$ = inspection time
 $x_2$ = preparation time
 $x_3$ = number of inspectors
 $x_4$ = experience level of inspectors
 $x_5$ = the complexity of the project measured using function point analysis in a logarithmic scale.

$$Y = \beta_0 + \beta_1 x_1 + \beta_2 x_2 + \beta_3 x_3 + \beta_4 x_4 + \beta_5 x_5 + e$$

*Equation E:*

Thus, with the system of MLR equations, a set of team coefficients is evaluated using: Equation F.

Evaluation of β coefficients is realized using Least Square Technique using Matlab support and requires a minimum of six empirical projects for the evaluation purpose [7]. Let Y represent IPM value for a project that can be obtained by substituting the parameter values and team coefficients in equation (E) as shown below.

Thus, with the system of MLR equations, a set of team coefficients is evaluated using:

Thus, having stabilized the team coefficients, the manager can obtain the desired IPM by appropriately tuning the inspection influencing parameters for the given complexity of the project [7]. Table 3 illustrates the computed IPM values for the previously sampled 15 projects.

Equation F:

$$[IPM] = [X][\beta] + [E] \text{ where}$$
$$[IPM] = [Parameters] \times [TeamCoefficients] + [Error Term]$$

$$M = \begin{bmatrix} IPM_1 \\ IPM_2 \\ IPM_3 \\ IPM_4 \\ - \\ - \\ IPM_n \end{bmatrix} \quad \beta = \begin{bmatrix} \beta_0 \\ \beta_1 \\ \beta_2 \\ \beta_3 \\ \beta_4 \\ \beta_5 \end{bmatrix} \quad X = \begin{bmatrix} 1 & x_{11} & x_{12} & x_{13} & x_{14} & x_{15} \\ 1 & x_{21} & x_{22} & x_{23} & x_{24} & x_{25} \\ 1 & x_{31} & x_{32} & x_{33} & x_{34} & x_{35} \\ 1 & x_{41} & x_{42} & x_{43} & x_{34} & x_{45} \\ - & - & - & - & - & - \\ - & - & - & - & - & - \\ 1 & x_{n1} & x_{n2} & x_{n3} & x_{n4} & x_{n5} \end{bmatrix} \quad E = \begin{bmatrix} e_1 \\ e_2 \\ e_3 \\ e_4 \\ - \\ - \\ e_n \end{bmatrix}$$

|  | P1 | P2 | P3 | P4 | P5 | P6 | P7 | P8 | P9 | P10 | P11 | P12 | P13 | P14 | P15 |
|---|---|---|---|---|---|---|---|---|---|---|---|---|---|---|---|
| Project hours(*) | 250 | 263 | 300 | 507 | 869 | 1806 | 2110 | 4248 | 4586 | 4644 | 6944 | 7087 | 7416 | 8940 | 9220 |
| IPM at req. phase | 4.57 | 5.40 | 6.89 | 5.71 | 2.714 | 3.11 | 1.25 | 0.66 | 0.36 | 1.03 | 0.35 | 0.66 | 0.51 | 0.73 | 0.60 |
| IPM at des. phase | 0.71 | 0.67 | 1.00 | 1.17 | 0.889 | 0.86 | 0.44 | 0.20 | 0.22 | 1.56 | 0.33 | 0.30 | 0.24 | 0.29 | 0.17 |
| IPM at imp. phase | 0.36 | 0.70 | 0.37 | 0.36 | 0.412 | 0.18 | 0.13 | 0.20 | 0.25 | 0.32 | 0.20 | 0.50 | 0.20 | 0.24 | 0.12 |
| Avg IPM | 1.88 | 2.25 | 2.75 | 2.41 | 1.34 | 1.39 | 0.61 | 0.35 | 0.27 | 0.97 | 0.29 | 0.49 | 0.32 | 0.42 | 0.29 |
| Tc (%) | 96.0 | 95.0 | 91.5 | 96.0 | 89.8 | 87.0 | 92.0 | 95.4 | 96.5 | 88.3 | 96.9 | 96.5 | 93.1 | 95.8 | 92.3 |

Table 3: IPM Estimation

Observable gains of IPM are:

1. IPM is introduced as an effort analysis metric in order to uniquely identify the effort put forth by the team for inspection.
2. The objective of introducing IPM as a quality indicator is to indicate the level of quality being achieved by the inspection team.
3. The main purpose of IPM is to provide transparency and visibility to the customers and thereby help them to justify and control the developmental cost.
4. Introduction of IPM in the software development cycle enables the inspection team to evaluate their performance level.
5. With the implementation of IPM as a software quality metric in the inspection process, the managers of the software company can now choose the team specification in order to achieve desired inspection effort.
6. The goal of IPM is to provide a deep visibility to the inspection performance for stakeholders, clients, managers, and outsourcing agents.
7. Existence of IPM in the software industry enables management to dynamically justify and control the staff cost to every project based on team performance.
8. IPM further acts as an awareness metric for the inspection-performing team to be aware of the team's performance and to appropriately formulate strategies towards their improvement activities.
9. Implementation of IPM therefore acts as a metric to save the economy of the company as it provides deep visibility of the team's performance in effective defect capturing abilities.
10. IPM further encourages the test team to train themselves for the capturing of residual defects.

DI and IPM are applicable across a variety of projects with the limitation that the ranges of values for acceptable performance differ from one type of project to another like innovative projects, legacy projects, etc.

### Conclusion

Inspection is one of the most promising techniques of defect management that enhances industrial productivity and quality. Inspection is a challenging task that provides a platform for a professional inspector to exhibit his competency to detect a maximum number of defects under time, cost, and resource constraints. It is essential for quality managers to apply appropriate metrics to monitor the effectiveness of inspection and the performance level of inspection. They can make use of two newly introduced metrics, DI and IPM, with their desirable band of operation to judge the level of success of process.

The investment of implementing DI and IPM over the existing process certainly demands a cost. However, it is inline with the dictum, "It is not just the investment that matters for quality, but also the right kind of investment." The implementation of DI and IPM is a right kind of investment that improves the position of a company's value to the market and stakeholders. The process metric DI and people metric IPM could be effectively used by clients, sponsors, and users to judge the perfection of developed, highly qualified software. Further, these metrics will pave the way for justifying the developmental cost with deep visibility into the process.

Due to the value of DI and IPM, the required effort to capture a maximum amount of defects is reduced. Managers get the added advantage of monitoring team performance, project after project, in a convincing way using numerical estimations through characteristic coefficients of the team or the company.

The DI and IPM value can now be either estimated based on defect counts from the shop floor or they can be predicted through the process coefficients and team coefficients that were empirically evaluated using a large sample of projects. Once the coefficients are stabilized, it is possible to predict the achievable DI and IPM through our model, without depending on the defect count. It implies that the managers can have the ability to finalize the inspection influencing parameters while planning the inspection process to achieve a particular DI. Having finalized the IPM that a company should achieve, it can tune the number of persons doing inspection, the experience of each person, and the time to be spent by each person to achieve the desired quality level of IPM.

Since DI and IPM are directly affecting defect management, development of a 99% defect-free product is possible by choosing appropriate values of parameters influencing DI and IPM. In order to realize the effectiveness of DI and IPM, it is absolutely necessary for quality-conscious outsourcing agencies and companies to run a piloted rollout of the inspection strategy.

DI and IPM (Nair-Suma metric) are valid across a spectrum of projects. But, it is important to note the ranges of values for acceptable performance differs from one type of project to another. ♦

### Acknowledgment:

The authors would like to acknowledge all the industry personnel who supported this work directly and indirectly within the framework of nondisclosure and thank all those who supported us to carry out the research successfully. Authors would further like to acknowledge the CrossTalk Editorial Board and the reviewers for their valuable suggestions.

### Disclaimer:

CMMI® is registered in the U.S. Patent and Trademark Office by Carnegie Mellon University.

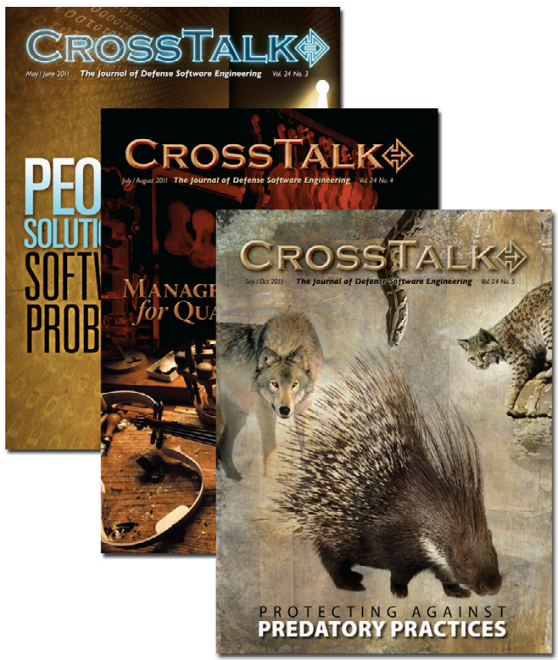

# CALL FOR ARTICLES

If your experience or research has produced information that could be useful to others, CrossTalk can get the word out. We are specifically looking for articles on software-related topics to supplement upcoming theme issues. Below is the submittal schedule for three areas of emphasis we are looking for:

**Rapid and Agile Stability**
*May/June 2012 Issue*
Submission Deadline: Dec 10, 2011

**The End of the PC**
*July/Aug 2012 Issue*
Submission Deadline: Feb 10, 2012

**Resilient Cyber Ecosystem**
*Sept/Oct 2012 Issue*
Submission Deadline: Apr 10, 2012

Please follow the Author Guidelines for CrossTalk, available on the Internet at <www.crosstalkonline.org/submission-guidelines>. We accept article submissions on software-related topics at any time, along with Letters to the Editor and BackTalk. To see a list of themes for upcoming issues or to learn more about the types of articles we're looking for visit <www.crosstalkonline.org/theme-calendar>.

## ABOUT THE AUTHORS

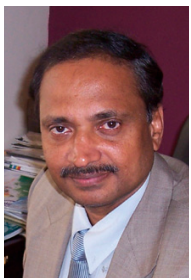

**Dr. T.R. Gopalakrishnan Nair** was the Vice President for Research in DS Institutions, Bangalore, India and currently holds the Saudi Aramco Endowed Chair of Technology in PMU, KSA. He worked in Indian Space Research and Indian Industries and has published extensively in computer field. Dr. Nair, a senior member of IEEE and ACM, was given the national technology award "PARAM" in 1992. He is interested in Software Engineering, Advanced Networking, Real-time Systems and Bio Computing.

T.R. Gopalakrishnan Nair
Saudi Aramco Endowed
Chair - Technology
Computer Engineering and Science
Prince Mohammad Bin Fahd University
PO Box 1664, Alkhobar 31952
Phone: +966 896 4554
Fax: + 91 896 4566
E-mail: trgnair@ieee.org
E-mail: trgnair@gmail.com

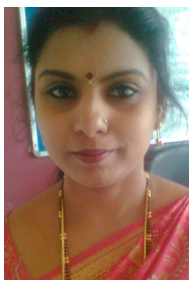

**Dr. V. Suma** is Head-Advanced Software Engineering Research Group of Research and Industry Centre and Professor in Dayananda Sagar Institutions, India. With a Ph.D. in Computer Science and Engineering, she is associated with several leading software industries, universities. Being an IEEE member, she has authored international book chapter and several research papers published at reputed International Journals, National, International conferences. Her main areas of interest are Software Engineering, Database Management System and Information Systems.

Suma. V.
Head, Advanced Software Engineering
Research Group,
Research and Industry
Incubation Center,
Professor, Department of Information
Science and Engineering
Dayananda Sagar Institutions,
Kumaraswamy Layout,
Bangalore-560078, India
Phone: +91 9448305148
Fax: +91 80 23637053
E-mail: Sumavdsce@gmail.com